\documentclass[aps,longbibliography,showpacs,twocolumn,superscriptaddress,amsmath,amssymb,verbatim]{revtex4-1}

\usepackage[thinlines]{easytable}
\usepackage{graphicx}
\usepackage{lmodern}
\usepackage{epstopdf}
\usepackage{dcolumn}
\usepackage{bm}
\usepackage{subfigure}
\usepackage{makecell}
\usepackage{amsmath} 
\usepackage[pdftex,colorlinks=true,citecolor=blue,linkcolor=blue,urlcolor=blue,bookmarks=true]{hyperref}

\usepackage{color}
\usepackage{textcomp}
\definecolor{red}{rgb}{1,0,0}

\definecolor{blue}{rgb}{0,0,1}

\definecolor{green}{rgb}{0,1,0}

\begin{document}
	\preprint{APS}

\title{Magnetic properties of triangular lattice antiferromagnets Ba$_{3}$\textit{R}B$_{9}$O$_{18}$ (\textit{R} = Yb, Er) 
 }

\author{J. Khatua}
\affiliation{Department of Physics, Indian Institute of Technology Madras, Chennai 600036, India}
\author{M. Pregelj}
\affiliation{Jo\v{z}ef Stefan Institute, Jamova cesta 39, 1000 Ljubljana, Slovenia}
\author{A. Elghandour}
\affiliation{Kirchhoff-Institut fürPhysik, Universität Heidelberg, INF 227, D-69120 Heidelberg, Germany}
\author{Z. Jagli\v{c}i\' {c}}
\affiliation{Institute of Mathematics, Physics and Mechanics, Jadranska 19, 1000 Ljubljana, Slovenia}
\affiliation{Faculty of Civil and Geodetic Engineering, University of Ljubljana, Jamova 2, 1000 Ljubljana, Slovenia}
\author{R. Klingeler}
\affiliation{Kirchhoff-Institut fürPhysik, Universität Heidelberg, INF 227, D-69120 Heidelberg, Germany}
\author{A. Zorko}
\affiliation{Jo\v{z}ef Stefan Institute, Jamova cesta 39, 1000 Ljubljana, Slovenia}
\affiliation{Faculty of Mathematics and Physics, University of Ljubljana, Jadranska u. 19, 1000 Ljubljana, Slovenia}
\author{P. Khuntia}
\email[]{pkhuntia@iitm.ac.in}
\affiliation{Department of Physics, Indian Institute of Technology Madras, Chennai 600036, India}
\affiliation{Quantum Centre for Diamond and Emergent Materials, Indian Institute of Technology Madras,
	Chennai 600036, India.}
\affiliation{Functional Oxide Research Group, Indian Institute of Technology Madras, Chennai 600036,
	India.}

\date{\today}

\begin{abstract}
	
	Frustration, spin correlations and interplay between competing degrees of freedom  are some of the key ingredients that underlie exotic states with fractional excitations  in quantum materials. Rare-earth based two dimensional magnetic lattice wherein crystal electric field, spin-orbit coupling, anisotropy and electron correlation between rare-earth  moments  offer a new paradigm in this context. 
	Herein, we present crystal structure, magnetic susceptibility and specific heat accompanied by crystal electric field calculations on the
	polycrystalline sample of Ba$_{3}$$R$B$_{9}$O$_{18}$ ($R$ = Yb, Er) in which $R^{3+}$ ions form a perfect triangular
	lattice without anti-site disorder. The localized $R^{3+}$ spins show neither long-range order
 nor spin-glass state down to 1.9 K in Ba$_{3}$$R$B$_{9}$O$_{18}$. 
Magnetization data reveal a pseudospin $J_{\rm eff}$ = 1/2 ( Yb$^{3+}$) in the Kramers doublet state  and a weak antiferromagnetic interaction between $J_{\rm eff}$ = 1/2 moments in the Yb variant.
 On the other hand,  the effective moment $\mu_{\rm eff}$ = 8.8 $\mu_{B}$ was obtained from the Curie-Weiss fit of the low-temperature susceptibility data of Er variant suggests the admixture of higher crystal electric field states with the ground state.
The Curie-Weiss fit of  low-temperature susceptibility data for Er system unveils the presence of a relatively strong antiferromagnetic interaction between Er$^{3+}$ moments compared to  its Yb$^{3+}$ analog. Ba$_{3}$ErB$_{9}$O$_{18}$ does not show long-range magnetic ordering down to 500 mK. Furthermore,  our crystal electric field calculations based on magnetization data of  Ba$_{3}$ErB$_{9}$O$_{18}$ suggest the presence of small gap between the ground and first excited Kramers doublets. The broad maximum around 4 K in magnetic specific heat in zero-field is attributed to the thermal population of
the first CEF excited state  in Ba$_{3}$ErB$_{9}$O$_{18}$, which is consistent with our CEF calculations. 	
\end{abstract}
\maketitle
\section{Introduction}
Geometrically frustrated magnets, wherein magnetic ions arranged on corner or side shared triangular motifs, have been of
intense research interest recently \cite{Balents2010,KHUNTIA2019165435}. In such materials, competing interactions accompanied by strong quantum fluctuations can melt long-range magnetic ordering leading to exotic
 ground states such as quantum spin-liquid (QSL) \cite{ANDERSON1973153,lacroix2011introduction}. 
Quantum spin liquids are characterized by the absence of phase transitions down to $T$ $\rightarrow 0 $  despite strong exchange interaction between spins. In the QSL state, spins maintain a highly entangled  state and support exotic fractional excitations that are essential ingredients for quantum computing \cite{ANDERSON1973153,lacroix2011introduction, Savary_2016,Broholmeaay0668,RevModPhys.80.1083}. 
Beyond the fundamental physics point of view, it is generally believed that high-$T_{c}$ superconducting state can be realized from the parent QSL state since the seminal proposal of P. W. Anderson in 1973 \cite{ANDERSON1973153,RevModPhys.78.17}. \\
The experimental realization of QSL state  in Cu$^{2+}$ ($S$ = 1/2) based frustrated magnet such as in triangular lattice $\kappa$-(ET)$_{2}$Cu$_{2}$(CN)$_{3}$ \cite{RevModPhys.89.025003}, kagomé lattice ZnCu$_{3}$(OH)$_{6}$Cl$_{2}$ \cite{Khuntia2020,Han2012} and hyperkagome lattice PbCuTe$_{2}$O$_{6}$ \cite{PhysRevLett.116.107203} due to frustration induced strong quantum fluctuations has  generated flurry of experimental and theoretical activities in  broadening our  understanding of exotic excitations in the entangled states  of  correlated  quantum matter \cite{Balents2010,RevModPhys.89.025003}. Especially, $S$ = 1/2 triangular lattice antiferromagnet is one of the simplest two-dimensional prototypical frustrated
quantum magnets that offer a versatile platform to realize remarkable quantum many-body phenomena, for example, QSL in 1T-TaS$_{2}$ \cite{Klanjsek2017} and Ba$_{3}$CuSb$_{2}$O$_{9}$ \cite{PhysRevLett.109.117203}, and quantum magnetization and continuum excitations in  Ba$_{3}$CoSb$_{2}$O$_{9}$ \cite{Ito2017} and unconventional spin dynamics in an hourglass magnet \cite{PhysRevB.87.214417}. Despite enormous efforts in past, the ideal realization of QSL  remains scarce due to anti-site disorder,  defects and presence of  complex magnetic interactions in real quantum materials \cite{PhysRevB.93.140408,PhysRevB.99.054412,PhysRevB.92.180411}.\\
Recently, rare-earth-based triangular lattice antiferromagnets in which anisotropic magnetic interactions induced by spin-orbit coupling and crystal electric field offer an alternate route to realize exotic quantum phenomena \cite{Arh2022}. Similar to 4$d$ and 5$d$ systems \cite{Takagi2019}, rare-earth magnet in which 4$f$ shells  are accommodated with odd number of electrons can also host a low energy effective-1/2 spin in the lowest Kramers doublet state.  For instance, rare-earth magnet YbMgGaO$_{4}$, where Yb$^{3+}$ ions form a triangular lattice in $a$$b$-plane show promising quantum many-body phenomena \cite{Li2015,https://doi.org/10.1002/qute.201900089}. The negative value of Curie-Weiss temperature ($\simeq -$ 4 K) obtained from the fit of low-temperature magnetic susceptibility data suggests the presence  of an  antiferromagnetic interaction between $J_{\rm eff}$ = 1/2 moments in  YbMgGaO$_{4}$.  A power law behavior of specific heat is attributed to a gapless quantum spin liquid state in  this Heisenberg antiferromagnet \cite{Li2015}. Muon-spin relaxation measurement reveals that localized  Yb$^{3+}$ spins in YbMgGaO$_{4}$ maintain a dynamic ground state  down to 60 mK \cite{PhysRevLett.117.097201}. Also, the presence of fractionalized spinon excitations was suggested by inelastic neutron scattering experiments \cite{Shen2016}.  
However, the presence of  disorder due to  Ga$^{3+}$/Mg$^{2+}$  site sharing  put  a strong constraint for the  unambiguous identification of ground state of YbMgGaO$_{4}$ \cite{PhysRevLett.120.087201,PhysRevLett.117.267202}.
Theoretically, it is suggested that the presence inter-site defects in the crystalline structure
provide an additional source to destabilize long-range magnetic order state by 
 the randomized exchange interactions \cite{PhysRevLett.119.157201}.\\ In this respect, rare-earth delafossite materials NaYbX$_{2}$ (X = O, S and Se) offer a
 unique opportunity to realize spin-orbit driven spin liquid state \cite{Liu_2018,PhysRevB.98.220409,PhysRevB.100.224417,Bordelon2019}. In delafossite materials, the nearest-neighbor Yb$^{3+}$ ions (3.34 {\AA}) are arranged on a triangular lattice without anti-site disorder between constituent atoms and the antiferromagnetic interaction between Yb$^{3+}$ moments is bit stronger compared to that in YbMgGaO$_{4}$. In addition, the thermodynamic and muon-spin resonance experiments on NaYb$X_{2}$ reveal that Yb$^{3+}$ spins do not undergo a magnetic long-range order down to 50 mK in zero-magnetic field \cite{Bordelon2019,PhysRevB.100.241116,PhysRevB.100.144432}.  However, most of the Yb based delafossite materials show  magnetization plateau and long-range magnetic order state in the presence of magnetic field that plays an important role in tuning the inter-plane interactions \cite{PhysRevB.98.220409, Bordelon2019,Bordelon2019}. The electron-spin resonance measurements on the single crystal of NaYbX$_{2}$ suggests the presence of anisotropic magnetic interactions between Yb$^{3+}$ spins \cite{Sichelschmidt_2019,doi:10.7566/JPSCP.30.011096}. More interestingly, the pressure induced Mott transition followed
 by the emergence of superconductivity has been observed in the QSL candidate NaYbSe$_{2}$ \cite{Jia_2020}.
 \begin{figure*}
 	\centering
 	\includegraphics[width=\textwidth]{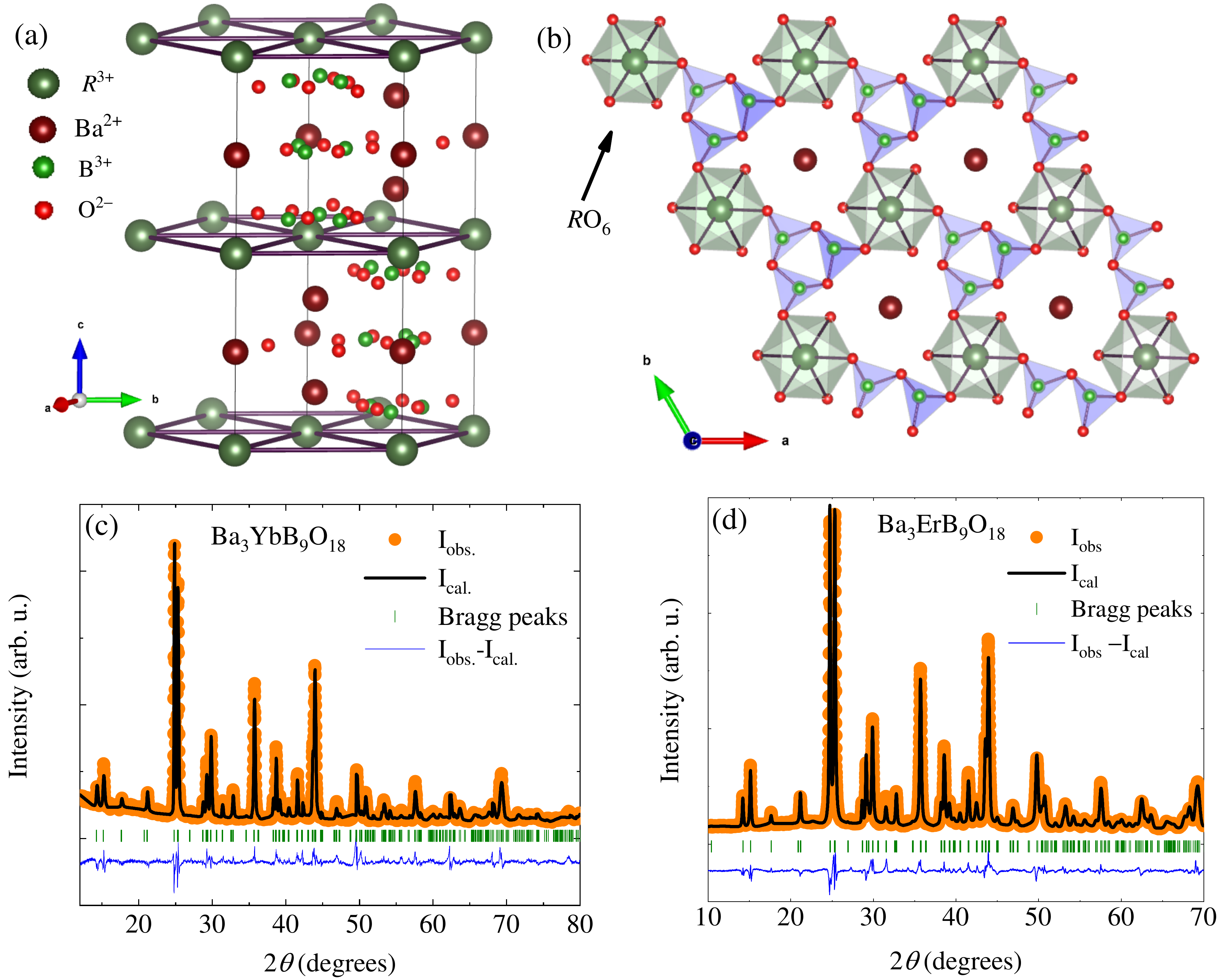}
 	\caption{(a) Schematic view of crystal structure of Ba$_{3}$$R$B$_{9}$O$_{18}$ ($R$ =Yb, Er) wherein solid lines denote the boundary of unit cell. $R^{3+}$ ions form a triangular lattice, which are stacked along $c$- axis with inter-planar distance 8.44 {\AA}. (b) $R$O$_{6}$ octahedra constituted  by nearest-neighbor oxygen ligand of $R^{3+}$ ions are shown.    (c) The Rietveld refinement of room temperature powder X-ray diffraction data of Ba$_{3}$YbB$_{9}$O$_{18}$. Experimentally observed points, the result of Rietveld fitting, expected Bragg reflection positions and difference between observed and calculated intensities are shown by orange circle,  black line, olive vertical bars, and blue line, respectively. (d) The Rietveld refinement of room temperature powder X-ray diffraction data of Ba$_{3}$ErB$_{9}$O$_{18}$.}{\label{xrd}}.
 \end{figure*}
  A similar scenario of pressure induced Mott transition and superconductivity is also observed in organic based  QSL candidate $\kappa$-(ET)$_{2}$Cu$_{2}$(CN)$_{3}$ \cite{PhysRevLett.117.107203}. In addition to next nearest-neighbor interaction, which melts 120$^{\circ}$ order \cite{PhysRevB.92.041105,PhysRevB.91.014426}, exchange anisotropy in Yb based triangular lattice antiferromagnets offer a novel route to realize spin-liquid state \cite{PhysRevB.94.035107,PhysRevB.95.165110}.\\
  Despite the large angular momentum, the Er (4$f$$^{11}$, $^{4}$I$_{15/2}$, $J$ = 15/2) member of the rare-earth series is also interesting to realize novel correlated quantum phenomena due to interplay between
   frustration and exchange anisotropy, which is a very fundamental requirement to understand quantum effects in rare-earth materials at low-temperature \cite{RevModPhys.82.53,PhysRevB.88.144402}.  In rare-earth magnetic  materials, the anisotropy originates from the combinations of spin-orbit coupling, local symmetry of rare-earth site and anisotropic super-exchange interaction. Similar to Yb system, the ground state of Er analog exhibits half-integer $J$ of Er$^{3+}$ ions due to crystal electric field  with different nature of exchange anisotropy  at low- temperature \cite{RevModPhys.82.53,PhysRevB.95.094422}.  Er-based magnets, where the $^{4}$I$_{15/2}$, multiplet splits in the local crystal electric-field environment into a Kramers doublet with the lowest energy for $J_{\rm eff}$ = 1/2, are ideal for hosting many anisotropy-driven ground states. For instance, the pyrochlore lattice Er$_{2}$Ti$_{2}$O$_{7}$ and triangular lattice K$_{3}$Er(VO$_{4}$)$_{2}$  exhibit  XY anisotropy \cite{PhysRevB.68.020401,PhysRevB.97.024415,PhysRevB.102.104423}  while the triangular lattice ErMgGaO$_{4}$ and hyperkagome lattice Er$_{3}$Ga$_{5}$O$_{12}$ show strong Ising anisotropy \cite{PhysRevB.101.094432,PhysRevB.100.184415,CEVALLOS20185}. Unlike ErMgGaO$_{4}$, the aforementioned Er magnets  undergo  a phase transition  at very low-temperature. ErMgGaO$_{4}$ does not order down to 25 mK, however,  it is not clear whether the randomness due to Mg$^{2+}$/Ga$^{3+}$ anti-site disorder leads to disordered ground state  in this antiferromagnet \cite{PhysRevLett.119.157201}. Recently, Er member of rare-earth delafossite  series AErCh$_{2}$ ($A$ = Na, K, Cs and Ch = S, Se, and Te)
   emerged as a structurally ideal platform  to explore spin-orbit driven quantum many-body phenomena \cite{PhysRevMaterials.3.114413,PhysRevB.103.144413,PhysRevB.102.024424,WeiweiLiu,PhysRevB.101.144432}. In the triangular lattice KErS$_{2}$, further neighbor in-plane magnetic interaction and easy-plane anisotropy stabilize antiferromagnetic ordered state below  0.2 K \cite{PhysRevB.103.144413}. Although the rare-earth delafossites exhibit magnetically ordered state at low-temperature, these systems are still of considerable research interest in recent years because it provides a rich reservoir to realize myriads of physical phenomena such as single-ion anisotropy, coexistence
   of three-dimensional and quasi-two-dimensional order, reduced moment
    in the ordered state, order by disorder state and transverse-field Ising model to name a few \cite{PhysRevLett.109.167201,PhysRevB.103.144413,PhysRevB.102.104423,Hester_2021}.          \\ The current interest is to investigate spin-orbit driven frustrated magnets in structurally perfect triangular lattice based on rare-earth ions and to explore the ground state properties under external perturbations in a controlled manner.\\ 
 Herein, we report crystal structure, magnetic susceptibility and specific heat results  on a  novel class of  rare-earth based magnet Ba$_{3}$$R$B$_{9}$O$_{18}$ ($R$ = Yb, Er), where $R^{3+}$ ions constitute a structurally perfect triangular lattice perpendicular to $c$-axis without anti-site disorder. In Ba$_{3}$$R$B$_{9}$O$_{18}$, the localized $R^{3+}$ ions interact antiferromagnetically albeit weak and exhibit neither long-range magnetic order  nor spin-glass state down to 1.9 K. Our results reveal a Kramers doublet state of Yb$^{3+}$ spin with an effective low energy state, $J_{\rm eff}$ = 1/2 at low temperature owing to crystal electric field  and spin-orbit coupling  for the Yb triangular lattice. Comparatively higher Curie-Weiss  temperature indicates the presence of a bit stronger antiferromagnetic interaction between Er$^{3+}$ moments in the triangular lattice Ba$_{3}$ErB$_{9}$O$_{18}$ compared to  its Yb analog. Furthermore, the Er variant does not show a phase transition down to 500 mK as revealed by our magnetization measurements. The presence of a broad maximum  around 4 K in zero-field specific heat data of Er triangular lattice suggests the presence  of low-lying crystal field excitations due to a small gap between two lowest Kramers doublet states. 
 \begin{figure*}
 	\centering
 	\includegraphics[width=\textwidth]{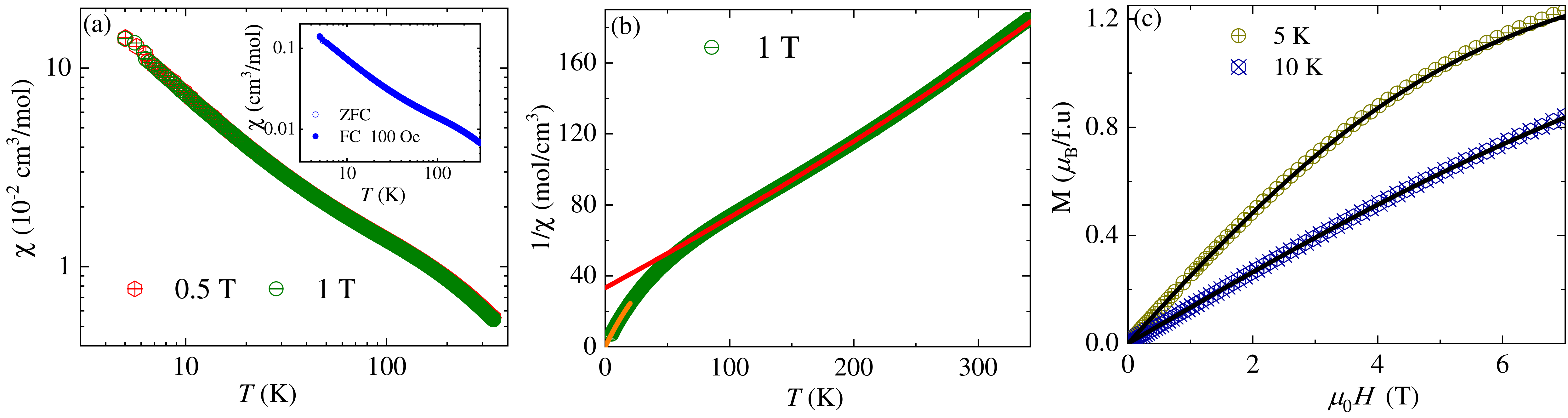}
 	\caption{(a) The temperature dependence of magnetic susceptibility $\chi(T)$ of BYBO in two different magnetic fields. The inset shows the temperature dependence zero-field cooled (ZFC) and field-cooled (FC)  magnetic susceptibility measured in $H$ = 100 Oe. (b) The temperature dependence of inverse  magnetic susceptibility. The red and orange lines are  Curie-Weiss fits to the  high-temperature and low-temperature inverse susceptibility data, respectively. (c) Magnetization as a function of external magnetic field at  5 K and 10 K and the solid lines are the Brillouin function fit of paramagnetic Yb$^{3+}$ ions.  }{\label{BYTO2}}.
 \end{figure*}
\section{Experimental details}  Polycrystalline samples of Ba$_{3}$$R$B$_{9}$O$_{18}$ (henceforth BYBO and BEBO) were prepared by a conventional solid state  method. BaCO$_{3}$ (Alfa Aesar, 99.997 \text{\%}), $R_{2}$O$_{3}$ (Alfa Aesar, 99.998 \text{\%})   and H$_{3}$BO$_{3}$ (Alfa Aesar, 98 \text{\%})  were mixed in stoichiometric quantities, while 10 \text{\%} excess H$_{3}$BO$_{3}$ was used due to its  volatile nature and  the reagent $R_{2}$O$_{3}$ was preheated at 700$^{\circ}$C for overnight to remove moisture  and carbonates prior to use.  The stoichiometric mixtures were pelletized and  the pellet was loaded into a platinum crucible for sintereing at 600$^\circ$C for 24 hours. This sintering process was performed at several intermediate temperatures and finally the single phase was obtained by annealing the sample at 950$^\circ$C for 48 hours. Powder X-ray diffraction (XRD) data of BRBO ($R$ = Yb, Er)  were collected  by employing Panalytical X$^{'}$pert PRO powder diffractometer with Cu $\mathcal{K\alpha}$ radiation ($\lambda $ = 1.54 {\AA}) at room temperature. Magnetization measurements were performed using a Quantum Design, SQUID VSM in the temperature range 5 K $\leq$ \textit{T} $\leq$ 340 K and in magnetic fields up to 7 T for BYBO samples. Magnetization measurements between 2 K $\leq$ \textit{T} $\leq$ 300 K in the applied magnetic field up to 5\,T were performed with Quantum Design MPMS XL-5 SQUID magnetometer using a closed-cycle cryostat for BEBO samples. Furthermore, low temperature magnetization measurements down to 0.5 K were carried out using the $^3$He option of MPMS, SQUID.
Fitting and modeling of the crystal-electric-field (CEF) effects was preformed using PHI software \cite{https://doi.org/10.1002/jcc.23234}.  Specific heat measurements were performed  using a Quantum Design, physical properties measurement system (PPMS) in the temperature range 1.9 K $\leq$ \textit{T} $\leq$ 250 K and in magnetic fields up to 7 T.\\
\section{results}
\begin{table}
	\caption{\label{tab:table1 }  Atomic coordinates of BYBO determined from the Rietveld refinement of X-ray diffraction data at 300 K. (Space group: P63/m, $ \alpha$ = $ \beta$ = 90.0$ ^{0}$, $\gamma$ = 120.0$ ^{0}$), $a$ = $b$ = 7.169 {\AA}, $c$ = 16.895 {\AA}
		and $\chi^{2}$ = 4.22, R$_{wp}$ = 5.88 \text{\%}, R$_{p}$ = 3.39 \text{\%}, and R$ _{exp}$ = 2.9 \text{\%})}
	
	\begin{tabular}{c c c c c  c c} 
		\hline \hline
		Atom & Wyckoff position & \textit{x} & \textit{y} &\textit{ z}& Occ.\\
		\hline 
		Yb & 2$b$ & 0 & 0 & 0 & 1 \\
		Ba$_{1}$ & 4$f$ & 0.666 & 0.333 & 0.1303 & 1 \\
		Ba$_{2}$ & 2$a$ & 0& 0 & 0.25 & 1 \\
		B$_{1}$ & 6$h$ & 0.518& $-$0.1544 & 0.25 & 1 \\
		B$_{2}$ & 12$i$ & $-$0.454& $-$0.314 & 0.076 & 1 \\
		O$_{1}$ & 6$h$ & 0.302& $-$0.165 & 0.25 & 1 \\
		O$_{2}$ & 12$i$ & 0.482& $-$.139 & 0.079 & 1 \\
		O$_{3}$ & 6$h$ & 0.653& 0.057 & 0.25 & 1 \\
		O$_{4}$ & 12$i$ & $-$0.283& $-$0.266 & 0.085 & 1 \\	
		\hline
	\end{tabular}
\end{table}
\begin{table}
	\caption{\label{tab:table2 }  Structural parameters of BEBO determined from the Rietveld refinement of X-ray diffraction data at 300 K. (Space group: P63/m, $ \alpha$ = $ \beta$ = 90.0$ ^{0}$, $\gamma$ = 120.0$ ^{0}$), $a$ = $b$ = 7.19 {\AA}, $c$ = 17.01 {\AA}
		and $\chi^{2}$ = 3.8, R$_{wp}$ = 6.5 \text{\%}, R$_{p}$ = 4.2 \text{\%}, and R$ _{exp}$ = 3.33 \text{\%})}
	
	\begin{tabular}{c c c c c  c c} 
		\hline \hline
		Atom & Wyckoff position & \textit{x} & \textit{y} &\textit{ z}& Occ.\\
		\hline 
		Er & 2$b$ & 0 & 0 & 0 & 1 \\
		Ba$_{1}$ & 4$f$ & 0.666 & 0.333 & 0.131 & 1 \\
		Ba$_{2}$ & 2$a$ & 0& 0 & 0.25 & 1 \\
		B$_{1}$ & 6$h$ & 0.503& $-$0.152 & 0.25 & 1 \\
		B$_{2}$ & 12$i$ & $-$0.463& $-$0.266 & 0.073 & 1 \\
		O$_{1}$ & 6$h$ & 0.292& $-$0.173 & 0.25 & 1 \\
		O$_{2}$ & 12$i$ & 0.504& $-$.117 & 0.083 & 1 \\
		O$_{3}$ & 6$h$ & 0.642& 0.041 & 0.25 & 1 \\
		O$_{4}$ & 12$i$ & $-$0.283& $-$0.271 & 0.089 & 1 \\	
		\hline
	\end{tabular}
\end{table}
\subsection{Rietveld refinement  and crystal structure of Ba$_{3}$\textit{R}B$_{9}$O$_{18}$ \textit{R} =(Yb, Er) } 
To confirm the phase purity, the  Rietveld refinement of X-ray diffraction data was performed using GSAS software \cite{doi:10.1107/S0021889801002242}. We used the atomic coordinates of Ba$_{3}$YB$_{9}$O$_{18}$ as the initial parameters to perform Rietveld refinement \cite{Li2004}.   The result of Rietveld refinement is shown in Fig.~\ref{xrd} (c) and (d) for Yb and Er variants, respectively. 
The obtained atomic parameters and goodness factors  are summarized in table \ref{tab:table1 } and \ref{tab:table2 }, which are in good agreement with earlier reports \cite{Li2004,cho2021studies}. Both the magnets, BYBO and BEBO, crystallizes in hexagonal structure with space group P63/m, where the $R^{3+}$ ions form a structurally perfect two-dimensional triangular lattice perpendicular to the $c$-axis, as shown in Fig.\ref{xrd} (a). The absence  of  anti-site disorder in these magnets  compared to well studied rare-earth triangular lattice  antiferromagnets  YbMgGaO$_{4}$ and ErMgGaO$_{4}$ suggests that the titled materials are promising candidates to explore quantum disordered states. The  magnetic $R^{3+}$  ion constitutes $R$O$_{6}$ octahedra with the nearest-neighbor oxygen ions. Furthermore, in-plane $R$O$_{6}$ octahedra with equal Yb-O distance (2.509 {\AA}) are connected through BO$_{3}$ triangle as shown in Fig.~\ref{xrd} (b).   One unit cell is composed of three triangular layers of $R^{3+}$ ions with inter-planar distance 8.44 {\AA} while the intra-plane distance between $R^{3+}$ ions is  the length of $a$-axis, which is associated with the type of rare-earth ion.   Structurally, the most striking difference of BYBO or BEBO with respect to triangular lattice YbMgGaO$_{4}$ and NaYbO$_{2}$ is that RO$_{6}$ octahedra are isolated in BRBO, whereas in YbMgGaO$_{4}$ and NaYbO$_{2}$, YbO$_{6}$ octahedra are connected with each other via a common oxygen ligand \cite{Li2015, Bordelon2019}.  This difference possibly modify the strength of magnetic exchange interaction between Yb$^{3+}$ moments as observed in the triangular lattice KBaYb(BO$_{3}$)$_{2}$ \cite{PhysRevB.103.104412}. In Ba$_{3}R$B$_{9}$O$_{18}$, we observed that the lattice parameters of Er triangular lattice are  bit higher than Yb triangular lattice. This obvious structural modification in BEBO is expected owing to a  different
ionic radius  of Er$^{3+}$ ion. Since there is no structural change, it is interesting to observe  the impact  of different $R^{3+}$ ions within the same triangular motif on the magnetic properties. 
\begin{figure*}[!]
	\centering
	\includegraphics[width=\textwidth]{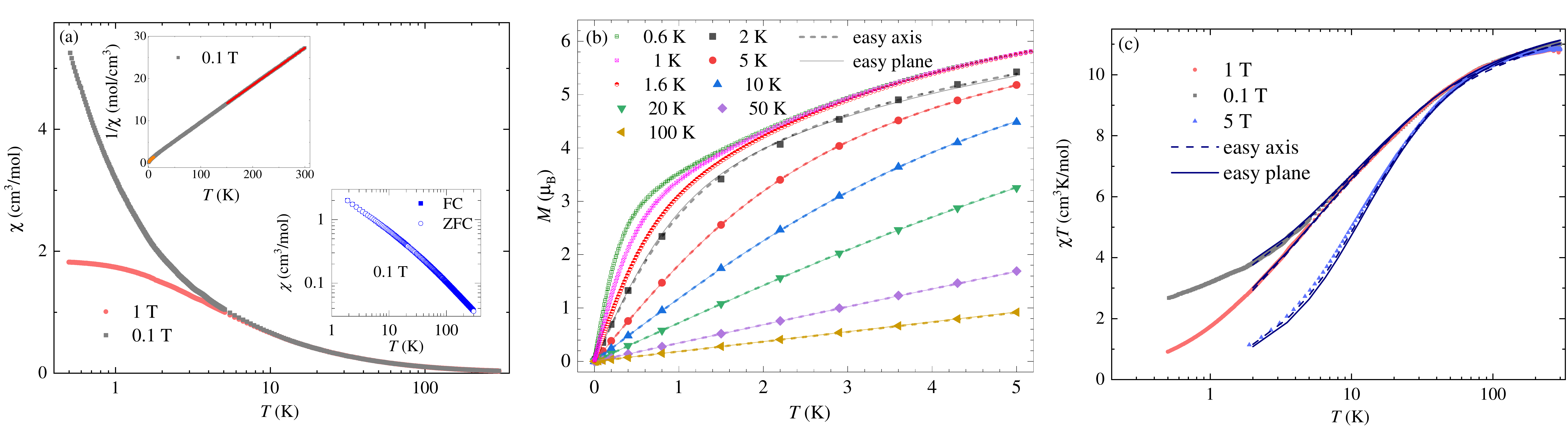}
	\caption{(a) The temperature dependence of the magnetic susceptibility $\chi$ of BEBO in different magnetic fields down to 500 mK. The top inset shows the temperature dependence inverse susceptibility where red and orange lines are the  Curie-Weiss fit to the high-temperature and low-temperature data, respectively. The right inset shows the temperature dependence of zero-field cooled (ZFC) and field-cooled (FC) magnetic susceptibility measured at 0.1\,T. 
			 (b) Magnetization as a function of external magnetic field at several temperatures. (c) The temperature dependence of $\chi T$ emphasizing the $\chi$ behavior at low temperatures. Lines in (b) and (c) correspond to the two CEF models presented in the text (only data for $T\geq5$\,K were fitted).}
	\label{fig-mag-BEBO} 
\end{figure*}
\begin{table*}[htb!]
	\centering
	\caption{Some rare-earth based frustrated triangular  lattice antiferromagnets with their  ground state properties.}.\\
	\begin{tabular}{ | c | c | c | c | c |c|c|c|c|}
		\hline
		$\makecell{\rm 	Materials \\ (\rm symmetry)}$ &   	$\makecell{\theta_{\rm CW} (\rm K)\\(\textnormal{ \rm high \textit{T}})}$   &	$\makecell{\theta_{\rm CW} (\rm K)\\(\textnormal{ \rm low \textit{T}})}$  & 	$\makecell{\mu_{\rm eff} (\mu_{B})\\(\textnormal{ \rm high \textit{T}})}$  & 	$\makecell{\mu_{\rm eff} (\mu_{B})\\(\textnormal{ \rm low \textit{T}})}$  & Anisotropy & $\makecell{\Delta (\rm CEF)\\ (\rm K) }$&$T_{\rm N} (\rm K)$ & Ref. \\[2 ex ] \hline
			\makecell{YbMgGaO$_{4}$ \\ ($R\bar{3}m$) } & - & $-$ 4  & - & 2.8 & Heisenberg & 440 & -  & \cite{Li2015,Paddison2017}   \\[2 ex] \hline
				\makecell{NaYbO$_{2}$ \\ ($R\bar{3}m$) } & $-$ 100  & $-$ 5.6  & 4.54 & 2.84 & XY & 404 & -  & \cite{PhysRevB.100.144432}   \\[2 ex] \hline
					\makecell{NaYbS$_{2}$ \\ ($R\bar{3}m$) } & $-$ 65 & $-$ 4.5  & 4.5 & 2.8 & XY & 266 & -  & \cite{PhysRevB.98.220409,PhysRevB.98.220409}   \\[2 ex] \hline
						\makecell{NaYbSe$_{2}$ \\ (($R\bar{3}m$)) } & $-$ 66 & $-$ 7  & 4.5  & 2.43 & XY & 183 & -  & \cite{PhysRevB.103.035144,PhysRevB.100.224417}   \\[2 ex] \hline
							\makecell{NaBaYb(BO$_{3}$)$_{2}$ \\ ($R\bar{3}m$1) } & $-$ 113 & $-$0.069  &  4.64 & 2.23 & - & - & -  & \cite{PhysRevMaterials.3.094404}   \\[2 ex] \hline
							\makecell{Rb$_{3}$Yb(PO$_{4}$)$_{2}$ \\ ($P\bar{3}m$1) } & $-$ 93 & $-$0.05  &  4.66 & 2.23 & - & - & -  & \cite{Guo2020}   \\[2 ex] \hline
								\makecell{NdTa$_{7}$O$_{19}$ \\ ($P\bar{3}m$1) } & $-$ 78 & $-$0.46  &  3.8 & 1.9 & Ising & 110 & -  & \cite{Arh2022}   \\[2 ex] \hline
								\makecell{ \textbf{Ba$_{3}$YbB$_{9}$O$_{18}$} \\ \textbf{($\textbf{P63/m}$)} } & $-$ \textbf{90} & $-$\textbf{0.12}  &   \textbf{4.73} & \textbf{2.32} & - & - & -  & \textbf{This work}  \\[2 ex] \hline
		\makecell{K$_{3}$Er(VO$_{4}$)$_{2}$ \\ (C2/c) } &  & $-$ 3  & - & - & XY & - & 0.15  & \cite{PhysRevB.102.104423}   \\[2 ex] \hline
		\makecell{NaErSe$_{2}$ \\ ($R\bar{3}m$) } & $-$ 10.9   & $-$4.3  & 9.5  & 9.4  & XY & - & - & \cite{PhysRevMaterials.3.114413,PhysRevB.102.024424,PhysRevB.101.144432}   \\[2 ex] \hline	
		\makecell{KErSe$_{2}$ \\ ($R\bar{3}m$) } & $-$ 8   & $-$ 3.8  & 9.5  & 9.4  & XY & 10.6 &0.2  & \cite{PhysRevB.103.144413,PhysRevMaterials.3.114413,PhysRevB.101.144432}   \\[2 ex] \hline
		\makecell{ErMgGaO$_{4}$ \\ ($R\bar{3}m$) } & $-$ 33   & $-$ 3.9  & 9.5  & 7  & Ising &-& - & \cite{CEVALLOS20185,PhysRevB.101.094432}   \\[2 ex] \hline
		\makecell{\textbf{Ba$_{3}$ErB$_{9}$O$_{18}$}  \\ ($\textbf{P63/m}$) } & $-$ \textbf{1}   & $-$ \textbf{0.5}  & \textbf{10.58}  & \textbf{8.8 } & - & 9.15 &-& \textbf{This work}   \\[2 ex] \hline			
	\end{tabular} 
 {\label{tableIN}}
\end{table*} 
\subsection{Magnetic susceptibility}
\subsubsection{\textnormal{ Ba$_{3}$YbB$_{9}$O$_{18}$}}
The temperature dependence of magnetic susceptibility of BYBO in magnetic fields  $\mu_{0}H$ = 0.5 T and 1 T is shown in Fig.~\ref{BYTO2} (a). The absence of anomaly in magnetic susceptibility  (see Fig.~\ref{BYTO2} (a))  suggests that Yb$^{3+}$ moments do not undergo a long-range magnetic ordering   down to 5 K. No zero-field cooled (ZFC) and field-cooled (FC) splitting  (see inset of Fig.~\ref{BYTO2} (a)) in 100 Oe rules out spin-freezing  in this magnet. The high-temperature susceptibility data follow  Curie-Weiss law, $\chi(T)$ = $\chi_{0}$ + $C$/($T-\theta_{\rm CW}$), where $\chi_{0}$ is the sum of  temperature independent core-diamagnetic and Van-Vleck susceptibility, $C$ is Curie-constant and $\theta_{\rm CW}$ is the Curie-Weiss temperature  representing a characteristic energy scale of interaction between magnetic moments of the materials under study. The Curie-Weiss fit (Fig.~\ref{BYTO2} (b)) of magnetic susceptibility data in the temperature range 100 K $\leq$ $T$ $\leq$ 340 K yields an effective moment, $\mu_{\rm eff}$ = $\sqrt{8C}$ = 4.73 $\mu_{B}$, which is comparable to that expected for  free Yb$^{3+}$ ions ( 4$f^{13}$, $^{2}$F$_{7/2}$), and $\theta_{\rm CW}$ = $-$ 90 K that is attributed to the presence of crystal-electric field. The nature of Yb$^{3+}$ spins  is expected to be different at lower-temperature due to interplay between spin-orbit coupling and crystal-electric field. 
In principle, the correlation between 4$f$ moments emerges at low-temperature. In order to get a rough idea about the energy scale of interaction between 4$f$ moments, magnetic susceptibility data were fitted with Curie-Weiss law in the temperature range 5 K $\leq$ $T$ $\leq$ 15 K.
The CW fit results $\theta_{\rm CW}$ = $-$ 0.12 $\pm$ 0.02 K and $\mu_{\rm eff}$ = 2.32 $\mu_{B}$. 
The obtained effective moment (2.32 $\mu_{B}$) is lower than the moment of free Yb$^{3+}$ ions according to Hund's rule, which suggests that the crystal-electric field leads to  Kramers doublet state of Yb$^{3+}$ spin with an effective low energy state, $J_{\rm eff}$ = 1/2.  The  small and negative value of $\theta_{\rm CW}$ indicates the presence of a weak  antiferromagnetic exchange interaction between Yb$^{3+}$  moments.\\      
The isotherm-magnetization of BYBO is depicted in Fig.~\ref{BYTO2} (c). The absence of finite magnetic moment in zero-field suggests that BYBO is free from  ferromagnetic signal. The observed field dependence  magnetization data are well captured (Fig.~\ref{BYTO2} (c)) by $M$/$M_{s}$ = $B_{1/2}$ ($y$),  where $B_{J}(y) = [\frac{2J+1}{2J} coth[\frac{y(2J+1)}{2J}]-\frac{1}{2J}coth\frac{y}{2J}]$ is the Brillouin function, $M_{s}$ (= g$J\mu_{B}$) is the saturation magnetization and $y=g\mu_{B}J \mu_{0}H/k_{B}T$, $\mu_{B}$ is the Bohr magneton, and $g$ is the Lande's g-factor. This fit yields powder average  Land\'e g factors  2.61 and  2.54 for 5 K and 10 K, respectively while $J$ was fixed to 1/2. The effective magnetic moment $\mu_{\rm eff}$ = 2.26 $\mu_{B}$ is obtained using $\mu_{\rm eff}$ =
$g\mu_{B} \sqrt{J(J + 1)}$, where $g$= 2.61 is known from the Brillouin
function fit. 
The obtained Land\'e g-factor  from the Brillouin fit leads to an effective moment for a low energy state, $J_{\rm eff}$ = 1/2,  which is close to that determined from the Curie-Weiss fit of $1/\chi$ data. The slight deviation from the Brillouin fit of magnetization isotherm at  5 K (Fig.~\ref{BYTO2} (c)) indicates the presence of a weak antiferromagnetic interaction between Yb$^{3+}$ ($J_{\rm eff}$ = 1/2) moments.
\subsubsection{\textnormal{ Ba$_{3}$ErB$_{9}$O$_{18}$}}
To observe the variation of magnetic properties by introducing a different rare-earth ion in the isostructural compound, we have measured temperature dependence of  magnetic susceptibility of BEBO in various magnetic fields.
Magnetic susceptibility ($\chi$) was measured in 0.1\,T in the ZFC and FC regime (inset of Fig.\,\ref{fig-mag-BEBO} (a)).
The two data sets exactly coincide, implying the absence of potential spin freezing.
The inverse susceptibility $1/\chi$ exhibits almost perfect linear temperature dependence, which can be fitted to the Curie-Weiss law yielding Curie-Weiss temperature $\theta$\,=\,$-$1.0(1)\,K for temperatures greater then 150\,K [in fact, $\theta$\,=\,$-$0.8(2)\,K all the way down to 20\,K].
The low-temperature Curie-Weiss fit yields (orange line in top inset of Fig.\,\ref{fig-mag-BEBO} (a))
 $\theta$\,=\,$-$0.45(10)\,K, implying a finite exchange interaction of the order of 0.5\,K.
\begin{table*}[!]
	\caption{CEF parameters (in cm$^{-1}$) for CEF Hamiltonian of Eq.\,(\ref{CEF-Ham}), $g$ factors of the ground state and sum of squares $\sum =\prod_{k=1}^{\text{data sets}} [\sum_{i=1}^{k~\text{points}}(x_{k,\text{exp}}-x_{k,\text{calc}})^2]$ for the two competing CEF models.
		\label{CEF-par}}
	\begin{ruledtabular}
		\begin{tabular}{c|c|c|c|c|c|c|c|c|c|c|c|c|c}
			model & $B_2^0$ & $B_4^0$ & $B_4^3$ & $B_4^{-3}$  & $B_6^0$  & $B_6^3$ & $B_6^{-3}$  & $B_6^6$ & $B_6^{-6}$ & $g_x$ & $g_y$ & $g_z$ & $\sum$ \\
			\hline
			easy axis & 1073 & 342.2 & 4313 & 1093 & 103.5 & -260.7 & -122.8 & 198.9 & 71.14 & 3.143 & 3.143 & 7.770 & 0.00824 \\
			\hline
			easy plane &-751.3 & 181.8 & 1525.7 & 773.5 & -6.867 & 71.23 & -226.0 & 265.1 & 913.0 &  2.086 & 6.575 & 6.575 & 0.0159 \\
		\end{tabular}
	\end{ruledtabular}
\end{table*}
In order to quantify the CEF effects we performed additional susceptibility measurements at higher fields, i.e., at 1 and 5\,T [Fig.\,\ref{fig-mag-BEBO}(c)], as well as magnetization measurements at several temperatures [Fig.\,\ref{fig-mag-BEBO}(b)].
Considering that the exchange interaction is probably of the order of 0.5\,K, we performed a combined fit of all the magnetization and susceptibility data for temperatures greater then 5\,K, i.e., approximately an order of magnitude greater then the expected exchange interaction.
According to the Hund's rule, the ground state multiplet of the Er$^{3+}$ ion is $^4I_{15/2}$, which is in a crystal field split into eight Kramers doublets composed of $|\pm m_J\rangle$ states [$m_J$\,=\,$(2n-1)/2$, where $n$\,=\,1-8].
The composition of the eight Kramers doublets directly depends on the CEF Hamiltonian, which can be written as
\begin{equation}
	H_{\text{CEF}} = \sum_{i,j} B_j^i O_j^i,
	\label{CEF-Ham}
\end{equation}
where $O_j^i$ are Stevens operators \cite{Stevens_1952} and $B_j^i$ are the corresponding scaling parameters. 
The relevant $B_j^i$ (Table\,\ref{CEF-par}) are determined by point symmetry at the Er$^{3+}$ site.
Indeed, we obtained a very good agreement with the data [lines in Fig.\,\ref{fig-mag-BEBO}(b) and (c)], but the fitting does not yield a unique solution.
In fact, two sets of CEF parameters (Table\,\ref{CEF-par}) both describe the data very well, yet they exhibit different magnetic ground states - an easy-axis and and easy-plane solution [dotted and solid lines in Fig.\,\ref{fig-mag-BEBO}(b) and (c), respectively].
Moreover, both models yield similar energy levels for the lower CEF states (Table\,\ref{CEF-level}), making both solutions rather difficult to distinguish.
Nevertheless, considering that the sum of squares \cite{https://doi.org/10.1002/jcc.23234} of the fit is by a factor of two smaller for the easy-axis solution, we favor this solution over the easy-plane one.
Finally, we note that small deviation of the calculated CEF responses at the lowest temperatures [Fig.\,\ref{fig-mag-BEBO}(b) and (c)] are likely due to small but finite exchange interaction.
We stress again that only data above 5\,K were fitted.
 Fig.~\ref{fig-mag-BEBO} (a) depicts the temperature dependence magnetic susceptibility of BEBO in different magnetic fields without any signature of a phase transition down to 500 mK. Below 2 K, isothermal magnetizations are shown in Fig.~\ref{fig-mag-BEBO} (b) as a function of magnetic fields up to 7 T. Future low temperature heat capacity and muon experiments below 500 mK may reveal interesting cooperative quantum phenomena in  this promising magnet.
\begin{figure*}
	\centering
	\includegraphics[width=\textwidth]{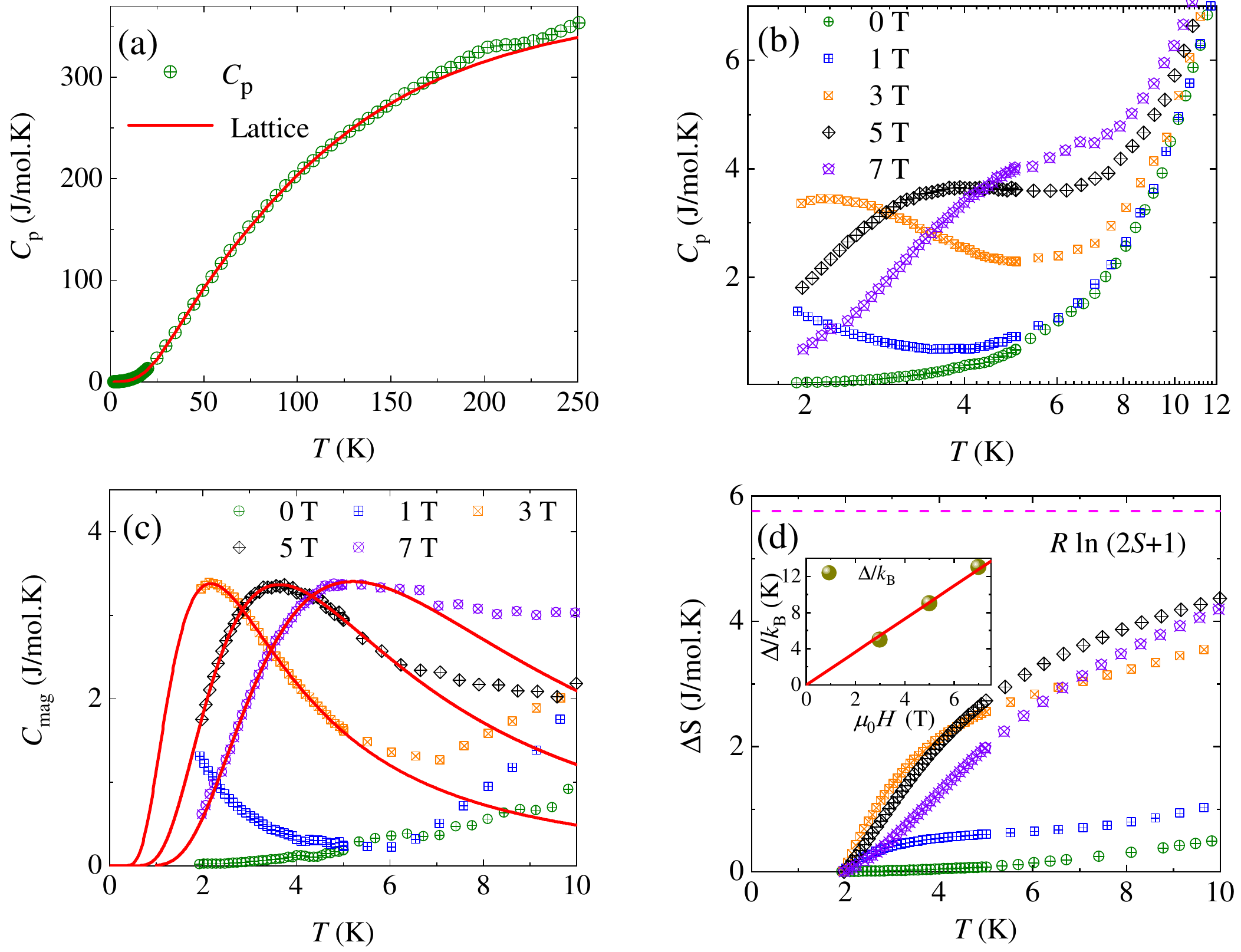}
	\caption{(a) The temperature dependence specific heat (\textit{C}$_{p}$) of BYBO in the temperature range 1.9 K $\leq$ \textit{T} $\leq$ 250 K in zero-magnetic field and the solid line is the fit to two Debye functions employed to extract the magnetic specific heat. (b) The
		temperature dependence of $Cp(T)/T$ in various magnetic fields. (c) The temperature dependence of $C_{\rm mag}(T)$ where solid red
		lines are the expected Schottky contribution i.e., Eq.~\ref{scho} due to Zeeman splitting of the lowest Kramers doublet state. (d) The
		entropy change ($\Delta S$ = $\int$ $C_{\rm mag}/T$ d$T$) as a function of temperature in the temperature range 1.9 K $\leq$ T $\leq$ 10 K in several
		magnetic fields up to 7 T and the dotted pink line is the expected entropy for $J_{\rm eff}$ = 1/2 moment. 
	 }{\label{BYTO3}}.
\end{figure*} 

\begin{figure*}
	\centering
	\includegraphics[width=\textwidth]{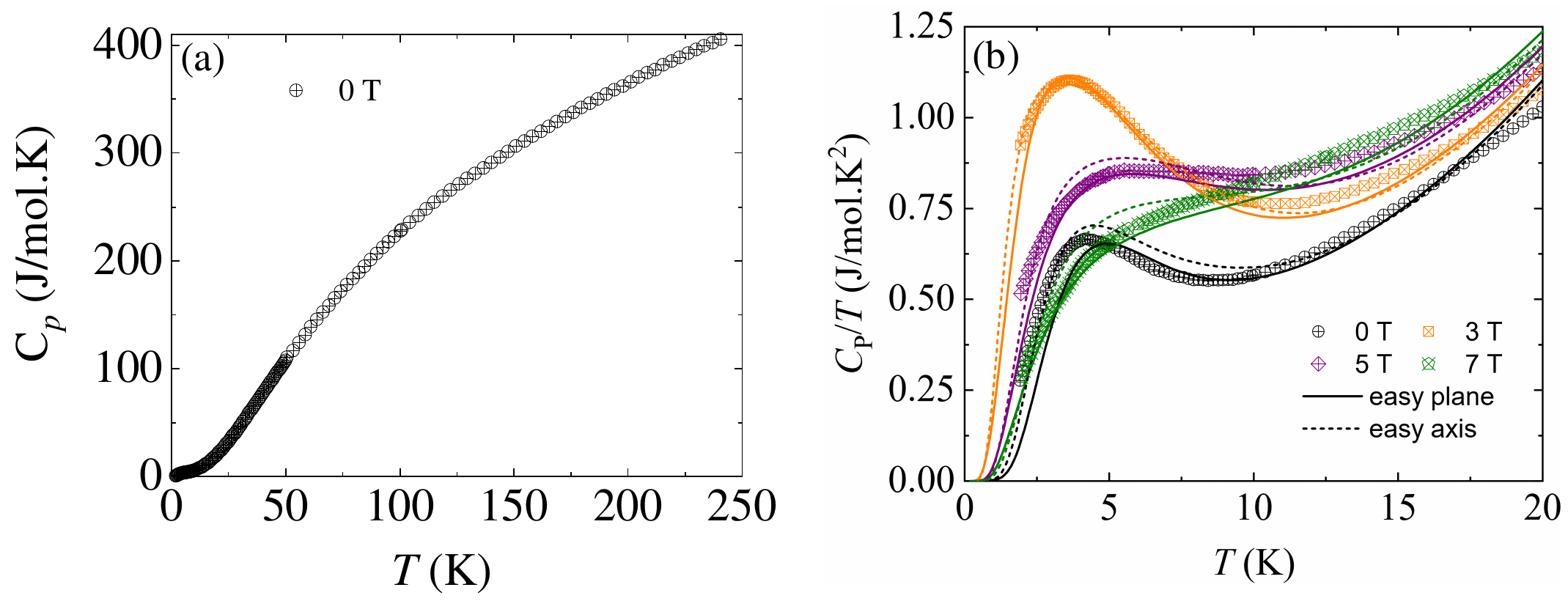}
	\caption{(a) The temperature dependence specific heat (\textit{C}$_{p}$) of BEBO in the temperature range 1.9 K $\leq$ \textit{T} $\leq$ 240 K in zero-field.  (b)  Temperature dependence of $C_{p}/T$ in  various magnetic fields while lines are the fitted curve by two CEF models and additional a $T^{3}$ term accounting for the lattice contribution as described in text. }{\label{BEBO2}}.
\end{figure*}
\subsection{Specific heat}
\subsubsection{\textnormal{ Ba$_{3}$YbB$_{9}$O$_{18}$}}
Specific heat studies are  ideal to understand the ground state properties of the titled rare-earth triangular lattice antiferromagnets. Fig.~\ref{BYTO3} (a) depicts the temperature dependence of total specific heat ($C_{p}(T)$) of BYBO measured in zero-field down to 1.9 K. The absence of an anomaly in specific heat indicates that Yb$^{3+}$ spins do not undergo a long-range magnetic ordering atleast in the measured temperature range. The measured specific heat is the sum of magnetic specific heat due to localized  Yb$^{3+}$ spins and lattice specific heat due to phonon contributions. The magnetic specific heat ($C_{\rm mag}(T)$) provides information concerning correlated magnetic 
\begin{table}[!]
	\caption{CEF energy levels (in cm$^{-1}$) for two competing CEF models.
		\label{CEF-level}}
	\begin{ruledtabular}
		\begin{tabular}{c|c|c|c|c|c|c|c|c}
			model & $E_0$ & $E_1$ & $E_2$ & $E_3$ & $E_4$ & $E_5$ & $E_6$ & $E_7$\\
			\hline
			easy axis & 0 & 9.15 & 25.2 & 74.3 & 391 & 395 & 457 & 1030 \\
			\hline
			easy plane &0 & 10.4 & 36.4 & 99.6 & 162 & 489 & 527 & 562\\
		\end{tabular}
	\end{ruledtabular}
\end{table} 
  phenomena and degeneracy of the ground state. Therefore, to extract magnetic specific heat, we consider Debye model i.e., $\textit{C}_{lat.}(\textit{T})= 9k_{\rm B}[\sum_{n=1}^{2}C_{n}(\frac{\textit{T}}{\theta_{D_{n}}})^{3}\int_{0}^{\theta_{D_{n}}/\textit{T}}\frac{x^{4}e^{x}}{(e^{x}-1)^{2}}dx]$ as lattice specific heat, where $k_{B}$ is the Boltzmann constant, $\theta_{D_{n}}$ (\textit{n} = 1, 2) are Debye temperatures and $C_{n}$ are the coefficients. As shown in Fig.~\ref{BYTO3} (a), in the temperature range 20 K $\leq$ $T$ $\leq$ 150 K, experimentally observed $C_{p}(T)$ data can be well reproduced by Debye model of lattice specific heat with $\theta_{D_{1}}$ = 230 K and $\theta_{D_{2}}$ = 345 K. In this fit, the  coefficients $C_{n}$ were fixed in the ratio $C_{1}$ : $C_{2}$ = 1:1.4, which is the ratio of total number of heavy atoms (Ba, Yb and B) and light atom (O) \cite{PhysRevX.9.031005,https://doi.org/10.1002/ejic.200500880}.
\\
In order to gain further insights into Kramers doublet state of Yb$^{3+}$ ions, specific heat measurements were performed in various magnetic fields and the results are  presented in Fig.~\ref{BYTO3} (b). It is observed that in the presence of magnetic field, the temperature dependence of  $C_{p}$ data exhibit a broad maximum. The specific heat data start increasing below 4 K in 1 T magnetic field, which probably suggests the onset of antiferromagnetic spin correlations and likely display  a broad maximum at much lower temperatures owing to weak exchange coupling between $J_{\rm eff}$ = 1/2 moments. The broad maximum shifts towards higher  temperature with increasing magnetic field. This scenario  is attributed to the Zeeman splitting of the lowest Kramers doublet state \cite{Li2015}.
\\
 The magnetic specific heat was obtained after subtracting the lattice specific heat from the total specific heat data and is shown in Fig.~\ref{BYTO3} (c). Next, we fitted high field  magnetic specific heat  data below 10 K to the two level Schottky anomaly \cite{PhysRevB.85.174438}
\begin{equation}
	C_{\rm sch}=f R \left(\frac{\Delta}{k_{B}T}\right)^2\frac{{\rm exp}(\Delta/k_{B}T)}{(1+{\rm exp}(\Delta/k_{B}T))^{2}},
	\label{scho}
\end{equation}
where $\Delta$ is the gap of Zeeman splitting of ground state Kramers doublet of Yb$^{3+}$ ion, $k_{B}$ is the Boltzmann constant, \textit{R} is universal  gas constant and \textit{f} measures the fraction of electron spin which contributes to the  splitting of ground state doublet. The best fit  was achieved with $\Delta/k_{B}$ = 4 K, 9 K and 13 K for magnetic field 3, 5 and 7 T, respectively as shown in Fig.~\ref{BYTO3} (c). Also, the estimated fraction of spin, $f$, is $\geq 93 $ \text{\%}, which illustrate the good reproducibility of
$C_{\rm mag}(T)$ data as this fraction is close to one. It is worth mentioning  here that this fraction does not measure the fraction of defect spins or orphan spins, which are normally found in the range of 5-10 \text{\%} due to defects in some disordered spin-lattices. In the titled material, the Schottky-like anomaly in the temperature dependence of magnetic specific  heat ($C_{\rm mag}$) is of magnetic origin due to  magnetic Yb$^{3+}$ ions. Inset of Fig.~\ref{BYTO3} (d) depicts the evolution of gap i.e., $\Delta/k_{\rm B}$ with magnetic field where red line is the linear fit which yields $g$ = 2.69 $\pm$ 0.04 \cite{PhysRevB.85.174438}.
This value of $g$ is close to that obtained from the analysis of magnetization data reflecting the consistency of our results and  accuracy of the fitting parameters obtained from two experiments. 
Fig.~\ref{BYTO3} (d) shows the temperature dependence of magnetic entropy change $\Delta S(T)$ = $\int C_{\rm mag}(T)/T \ \ dT$ in various magnetic fields.  It is observed  that $\Delta S$
tends to saturate just below the expected value (pink dashes line) of Yb$^{3+}$ spin for $J_{\rm eff}$ = 1/2. As the broad maximum for zero magnetic field is not
reached above 2 K we recover the entropy change much lower than $R$ln2 expected for
a Kramers doublet in the ground state. This suggests the presence of intrinsic magnetic specific heat at low-temperature. 
\subsubsection{\textnormal{ Ba$_{3}$ErB$_{9}$O$_{18}$}}
Fig.~\ref{BEBO2} (a) represents the specific heat $C_{p}(T)$ of BEBO in zero-field down to 1.9 K. For the clear visualization of low-temperature heat capacity data, the temperature dependence of $C_{p}/T$ in several magnetic fields is shown in Fig.~\ref{BEBO2} (b). It is apparent that there is no sharp anomaly down to 1.9 K suggesting the absence of  long-range antiferromagnetic order of Er$^{3+}$ moments. Rather, the zero-field specific heat data exhibit a broad maximum  $T_{\rm max}$ $\approx$ 4 K  suggesting the presence of low-lying crystal field excitations, which  is also observed in several rare-earth based frustrated  triangular lattice  antiferromagnets \cite{https://doi.org/10.48550/arxiv.2108.09693,PhysRevMaterials.3.114413}.\\ Indeed, the sum of magnetic-heat contribution calculated for the CEF parameters in TABLE \ref{CEF-par} (using program PHI), $C_{\rm mag}$, and phonon contribution that is at low temperatures proportional to $T^3$, i.e., $C_p = C_{\rm mag} + aT^3$, describes the observed behavior rather well [lines in Fig.~\ref{BEBO2}(b)]. We note that $C_{\rm mag}$ was scaled by ~0.85 to get a better agreement with the experiment. Still, the two models yield similar results, making it rather difficult to choose a favorable one. Nevertheless, the good agreement with the experiment suggests that the magnetic specific heat due to exchange interactions  between Er$^{3+}$ spins is expected at low-temperature atleast below 1.9 K \cite{PhysRevB.62.6496,Xing2020}.     
\section{Discussion}
Frustrated triangular lattice antiferromagnets offer an emblematic model for the experimental realization  of novel quantum states and exotic excitations in quantum materials.
Rare-earth magnets where 4\textit{f } shells are occupied with odd number of electrons can host effective 1/2 moments  which combines
spin and orbital moments in the Kramers doublet state.
One may expect such $J_{\rm eff}$ = 1/2 moment of Yb$^{3+}$ ion in BYBO (Yb$^{3+}$, 4$f^{13}$) wherein crystal electric field  splits the spin-orbit induced $2J+1$ = 8  degenerate ground state ($^{2}{J}_{7/2}$) into four Kramers doublets. 
Indeed, the estimated effective moment $\mu_{\rm eff}$ = 2.32 $\mu_{B}$ obtained from magnetic susceptibility data
suggests the  formation of Kramers doublet state in BYBO below 10 K. The low and negative Curie-Weiss temperature from the fit of the low-temperature susceptibility data indicates the presence of a weak antiferromagnetic interaction between $J_{\rm eff}$ = 1/2 moments of Yb$^{3+}$ ion. In rare-earth magnets, the magnetic interaction is mainly governed by dipolar and super-exchange interaction despite strong localization of 4$f$ electrons. For instance, in YbMgGaO$_{4}$, the  nearest-neighbor dipolar interaction between intra-layer (3.4 {\AA}) Yb$^{3+}$ moments is of the order of 0.16 K, which is 11 \text{\%} of the nearest-neighbor exchange interaction 1.36 K as estimated from Curie-Weiss temperature \cite{https://doi.org/10.1002/qute.201900089}.  This suggests that dominant exchange energy in YbMgGaO$_{4}$ is due to antiferromagnetic super-exchange interaction via Yb$^{3+}$-O$^{2-}$-Yb$^{3+}$ virtual electron hopping processes  \cite{PhysRevB.83.094411}.
 Whereas, in BYBO, the intra-planar distance  of Yb$^{3+}$ ion is 7.16 {\AA} that is almost double than the intra-planar distance in YbMgGaO$_{4}$, which suggests the presence of weak magnetic dipole–dipole interaction. The calculated dipole–dipole interaction in BYBO, following a phenomenological expression, E$_{\rm dip}$$ \approx$ $\mu_{0}g_{\rm avg}^{2}\mu_{B}^{2}/4\pi a^{3}$ (where $g_{\rm avg}$ is the powder average Land\'e $g$ factor and $a$ is the nearest-neighbor Yb-Yb distance) is $\sim$  0.011 K, which is 28 \text{\%} of the nearest-neighbor exchange interaction 0.04 K as estimated from Curie-Weiss temperature \cite{PhysRevLett.73.3306,https://doi.org/10.1002/qute.201900089}. Therefore, in BYBO, super-exchange interaction is weak as expected for the presence of isolated YbO$_{6}$ octahedra that connects the intra-plane YbO$_{6}$ octaedra via virtual exchange path Yb-B-O-B-Yb. A similar scenario of low exchange interaction is observed in another rare-earth based triangular lattice KBaYb(BO$_{3}$)$_{2}$, where intra-plane YbO$_{6}$ octahedra are not connected via a common O$^{2-}$ ions rather separated
 spatially within the $ab$-plane by BO$_{3}$ triangles \cite{PhysRevB.103.104412}. In view of this, it is suggested  that super-exchange interaction is the main mechanism for magnetic coupling in the present  Yb based triangular lattice \cite{PhysRevB.98.054408}. 
The specific heat data reveal the absence of magnetic ordering down to 1.9 K and in high magnetic field, specific heat data exhibit broad maximum which shifts to higher temperatures
with increasing magnetic field. The shift of broad maximum in $C_{m}$ is attributed  to the Zeeman splitting of the lowest Kramers doublet state of BYBO as observed in a few frustrated rare-earth  magnets. It is important to note here that the gap is proportional to the applied field and it yields $g$ factor that is in agreement with susceptibility. It is noted that we could not recover magnetic entropy close to $R$ ln 2 for $J_{\rm eff}$ = 1/2 spin state, because we did not take into account intrinsic  magnetic specific heat below 1.9 K due to limitation of our  instrument.  \\
In Ba$_{3}$ErB$_{9}$O$_{18}$, the reduction of Er$^{3+}$ moment (Er$^{3+}$, 4$f^{11}$) suggests the local symmetry of Er$^{3+}$ could split 2$J$+1 ($J$ = 15/2) multiplet into  Kramers
doublets.   
Interestingly, we obtained a comparably higher Curie-Weiss temperature ($-$0.5 K), which suggests the presence of antiferromagnetic  super-exchange interaction  between Er$^{3+}$ moments in BEBO. In contrast to BYBO, we observed a broad maximum in specific heat data of BEBO that is attributed to crystal field excitations.   This is further supported by our CEF calculations. Although varieties of rare-earth ions stabilize the same crystal symmetry, their magnetic interaction strength and anisotropy are different, for  instance,  rare-earth pyrochlore \cite{annurev-conmatphys-031016-025218} and the rare-earth
delafossites \cite{Bordelon2019,PhysRevB.103.144413}. A similar scenario is also observed in the present antiferromagnets that occur due to slight differences in bond length and dipolar interactions. In BEBO, the dipolar interaction between Er$^{3+}$ ions is expected to be larger  due to larger moment of Er$^{3+}$ (8.8 $\mu_{B}$) ions. In addition, the exchange interaction is associated with the nature of angular momentum states in the ground state \cite{PhysRevB.98.054408}. Therefore, one can expect different strength of exchange interaction due to different anisotropy between BEBO and BYBO. 
Low-temperature thermodynamic and local probe measurements in Ba$_{3}$RB$_{9}$O$_{18}$ are essential   to understand nature of magnetic anisotropy. It is worthwhile to explore the role of rare-earth magnetic ions  and anisotropy on the underlying magnetism and spin dynamics of this new family of rare-earth
based triangular lattice antiferromagnets. This may be highly relevant for establishing paradigmatic theoretical models in understanding a large class of frustrated materials wherein exotic ground state properties borne out of the intertwining of crystal electric field, spin-orbit coupling and electron correlations.    
\section{Conclusion}
In summary, we have synthesized and carried out magnetization and specific heat studies on a new series of rare-earth based magnets Ba$_{3}$$R$B$_{9}$O$_{18}$ ($R$ = Yb, Er).
    The present family of rare-earth based compound represents a structurally perfect magnet in which $R^{3+}$ ions constitute  a two dimensional triangular lattice without anti-site disorder.  The investigated Er based triangular lattice antiferromagnets neither exhibit signature of  long-range magnetic ordering nor spin-glass behavior down to 500 mK.   Magnetic susceptibility data of Yb member of the series reveal  the presence of  a  Kramers doublet of Yb$^{3+}$ ions with  an effective low energy  $J_{\rm eff}$ = 1/2 state  and a weak antiferromagnetic interaction between $J_{\rm eff}$ = 1/2 moments. We found a broad maximum around 4 K in zero-field  specific heat  in  the Er analog BEBO due to crystal field excitations,  which is consistent with  the presence of relatively small gap between two lowest Kramers doublet sate  as reflected in our CEF calculations. The characteristic energy scale of interaction  between Er$^{3+}$ moments is weak, which is typical for rare-earth based  frustrated triangular lattice antiferromagnets YbMgGaO$_{4}$ and ErMgGaO$_{4}$ to name a few. The present class of rare-earth triangular lattice antiferromagnet offer a promising platform to  realize frustration driven quantum phenomena given the two dimensional  nature of spin-lattice and the presence of spin-orbit interaction. Microscopic experimental techniques such as neutron scattering and muon spin relaxation  experiments and  theoretical calculations are desired to explore  the ground state properties  and associated excitations driven by spin-orbit interactions, crystal electric field and electron correlations in this promising class of  rare-earth based triangular lattice antiferromagnets.
    \section{Acknowledgments}
     PK acknowledges the funding by the Science and
    Engineering Research Board, and Department of Science and Technology, India through Research Grants.
\bibliographystyle{apsrev4-1}
\bibliography{BRBO}
\end{document}